# Flexible and Transparent All-Graphene Circuits for Quaternary Digital Modulations

Seunghyun Lee[1], Kyunghoon Lee[1], Chang-Hua Liu[1], Girish S. Kulkarni[1] and Zhaohui Zhong[1]*

[1] *Department of Electrical Engineering and Computer Science, University of Michigan, Ann Arbor, MI 48109, USA.*

* e-mail: zzhong@umich.edu

**In modern communication system, modulation is a key function that embeds the baseband signal (information) into a carrier wave so that it can be successfully broadcasted through a medium such as air or cables. Here we report a fully bendable all-graphene modulator circuit with the capability to encode a carrier signal with quaternary digital information for the first time. By exploiting the ambipolarity and the nonlinearity in a graphene transistor, we demonstrated two types of quaternary modulation schemes: quaternary amplitude-shift keying and quadrature phase-shift keying. Remarkably, both modulation schemes can be realized with just 1 and 2 all-graphene transistors, respectively, representing a drastic reduction in circuit complexity when compared with conventional modulators. In addition, the circuit is not only flexible but also highly transparent (~95% transmittance) owing to their all-graphene design with every component (channel, interconnects, load resistor, and source/drain/gate electrodes) fabricated from graphene films.**

Physically compliant electronics with the capability to conform to a non-planar surface is a field of rapidly growing interest due to the numerous possibilities it offers. Applications ranging from flexible solar cells [1], displays [2], e-papers [3], wearable electronics [4], and biomedical skin-like devices [5,6] open up new opportunities in the field of electronics. However, nearly all flexible electronic devices require an external power supply and data communication modules, and the lack of portability can severely limit the functionality of various applications. To drive the field forward, three key challenges need to be addressed: a means to generate or store power (e.g., flexible batteries or power generators), a data collecting scheme (e.g., flexible sensors), and a system to transmit and receive the collected data (e.g., flexible wireless communication scheme). Recent advances in the field have led to notable progress in the two areas of flexible power [7,8] and flexible sensors [5,6]. However, designing and manufacturing a flexible wireless communication system is still a challenge due to material constraints. Conventional organic polymers [9], amorphous silicon [10], or oxide-based thin film transistors [11,12] show only modest performance in this area owing to their limited carrier mobilities.

In this regard, graphene is the ideal material for flexible high speed communication systems due to its unique electronic and physical properties, including high carrier mobility [13], ambipolarity [14,15], transparency [16], and mechanical flexibility [17]. For example, graphene transistors have achieved unity gain cut-off frequencies of up to 300 GHz [18,19]. Graphene was also used as the channel and the gate material for flexible transistors owing to its mechanical flexibility [20,21]. Despite graphene transistors' low on/off ratio which limits their usage in the digital/logic applications, they are attractive in the analogue/radio frequency applications due to graphene's extremely thin structure that allows shorter scaling of channel length without the adverse short channel effects [22]. Beyond individual transistor, pioneering works on graphene analogue electronics led to the demonstration of graphene-based frequency doublers [23,24,25], mixers [26,27], and modulators [24,28,29] on rigid substrates. Graphene mixers were shown to effectively suppress odd-order intermodulations by exploiting the symmetric character of graphene transistors [26]. A high-performance mixer fabricated by integration of graphene transistors and passive components on a single silicon carbide wafer was also demonstrated [27]. Taking it one step further, several groups demonstrated binary digital modulation schemes [binary phase-shift keying (BPSK) and binary frequency-shift keying (BFSK)] with graphene transistors on rigid substrates [24,28,29]. Despite the remarkable progress, all previous graphene analog circuits were demonstrated on rigid silicon substrates. A flexible digital modulation scheme, which is the key building block for flexible high speed data communication, has not been realized using graphene circuits. In addition, previous graphene binary modulators can only encode single bit of data per symbol.

To this end, we demonstrate, for the first time, flexible and transparent all-graphene circuits for quaternary digital modulations that can encode two bits of information per symbol. The entire circuits are both flexible and transparent with every part of the circuit—including the transistor channels, the interconnects between transistors, the load resistance, and the source/drain/gate electrodes—fabricated with graphene only. The monolithic structure allows unprecedented mechanical flexibility and near-complete transparency which is not possible with either silicon or metal. This structure is possible due to graphene's unique property of being a zero-bandgap material retaining the property of both metal and semiconductor. Importantly, the ambipolarity of graphene transistors drastically reduces the circuit complexity when compared with silicon-based modulators. No more than a couple of transistors are required for the two quaternary modulation schemes demonstrated, whereas a multitude of transistors are required for conventional modulator circuits [30,31].

**Results**

**Modulation mechanism and transistor characteristics.**

The basic modulation techniques map the information by varying up to three different parameters (amplitude, frequency, and phase) of the carrier wave to represent the data. The most fundamental binary digital modulation techniques that correspond to each of these three parameters are binary amplitude-shift keying (BASK), binary frequency-shift keying (BFSK), and binary phase-shift keying (BPSK). Furthermore, by combining two or more binary modulation schemes, it is possible to extend this technique into quaternary digital modulation schemes such as quaternary amplitude-shift keying (4-ASK) and quadrature phase-shift keying (QPSK) [32]. Specifically, QPSK explores all four quadrants of the constellation, and it is the key building unit for highly efficient modulation techniques that are widely used in today's telecommunication standards such as Code division multiple access (CDMA) and Long term evolution (LTE). The above-mentioned binary and quaternary digital modulation schemes are plotted in a polar constellation with the radial coordinate as the amplitude and the angular coordinate as the phase (Fig. 1a). Importantly, all of them can be realized by using all-graphene circuits on the flexible and transparent platform.

Figure 1b shows the transmittance value of the graphene circuit on top of a plastic substrate (polyethylene naphthalate) as a function of the light wavelength (see Supplementary Methods for details of transmittance measurement). Fig. 1b (inset) shows the structure of the device fabricated on a bendable plastic substrate. The top, middle, and bottom graphene layers form the top gate layer, the channel/interconnect layer, and the bottom gate layer, respectively. Two dielectric layers are deposited using an atomic layer deposition method to isolate the three graphene layers. The final device is highly transparent as shown in Fig. 1b (~95% transmittance at 550nm wavelength) and fully bendable as shown in Fig. 1c. Although three layers of graphene were transferred, the overall transmittance is higher than the expected value of 93% [16] because the majority of the area is covered with only one layer of graphene after patterning and two layers of $Al_2O_3$. Only the channel area which occupies little space would have all three graphene layers (the bottom gate, the channel, and the top gate) overlapping each other. Under the optical microscope, the graphene devices can be identified by the contrast difference among the top gate, channel, and bottom gate region of the all-graphene transistor (Fig. 1c inset). The gate response curve was measured for each all-graphene transistors, and the yield was over 98% with 64 out of 65 transistors being fully functional. Figure 1d (inset) shows a typical gate response curve from the fabricated all-graphene transistors. Slight shift in the charge neutrality point is observed due to environmental doping. The carrier mobility value can be extracted by fitting the experimental value of source-to-drain conductance over varying gate voltages [33]. The device presented in Fig. 1d (inset) has a hole carrier mobility value of 3342±26 cm$^2$/Vs and electron carrier mobility of 2813±11 cm$^2$/Vs (See Supplementary Fig. S1 and Supplementary Methods). Figure 1d also presents a histogram for hole mobility values extracted from 30 different samples. The average hole mobility is 1771 cm$^2$/Vs with a standard deviation of 983 cm$^2$/Vs. These mobility values are several orders of magnitude higher than those of alternative materials such as organic polymer [9] and amorphous materials [10] as expected. More importantly, the unique ambipolar gate response of graphene transistors allows simple implementation of previously mentioned binary modulation schemes as illustrated in Fig. 1e. The amplitude, frequency, or phase of the output voltage will be determined by the operating gate bias point of the graphene transistor. For example, amplitude modulation (AM) can be achieved by utilizing the transconductance change over the gate voltage difference. Frequency modulation (FM) is achieved by interchanging the bias point from a region dominated by electron (or hole) carriers to the charge neutrality point. Similarly, phase modulation (PM) is realized by changing the bias point from an electron (or hole) carrier dominated region to the hole (or electron) carrier dominated region.

**Binary and quaternary modulation with a single transistor.**

We next demonstrated the three binary modulation schemes by using the all-graphene circuit. Figure 2a shows the circuit diagram overlaid on a false colour image of the device. Green, grey, and red are the respective colours for the top, middle, and bottom graphene layers. A graphene transistor was used for the modulation and another unbiased graphene transistor was used as the load resistor ($R_L$) for output ($V_{out}$). The middle graphene layer (grey) serves as the transistor channel, the interconnect between the transistor, the load resistor, and the source/drain electrodes. To achieve digital modulation, the carrier wave ($V_{\text{carrier}}$), the data bitstream ($V_{\text{data}}$), and the DC gate bias ($V_{\text{gs}}$) are added together and applied to the top gate (green) of the modulating transistor. Both $V_{\text{gs}}$ and $V_{\text{data}}$ determine the operating bias point of the transistor and modulate the carrier signal accordingly. The bottom gate (red) delivers additional flexibility to the measurement, and it can also be used to adjust the charge neutrality point ($V_{\text{Dirac}}$) if there is environmental doping.

Figure 2b,c,d shows plots of three basic binary modulation schemes demonstrated with the all-graphene circuit. For BASK, the sum of $V_{\text{data}}$ and $V_{\text{carrier}}$ is superimposed on $V_{\text{gs}}$ so that the different transconductances on different bias points will allow $V_{\text{carrier}}$ to change in amplitude at the output . Binary information of 0 and 1 is successfully represented by the low and high amplitude of carrier signal, respectively (Fig. 2b). Similarly, we control $V_{\text{gs}}$ and $V_{\text{data}}$ to adjust the bias point for both BFSK and BPSK. 0 and 1 are successfully differentiated by the doubling in frequencies (BFSK, Fig. 2c), or by the 180° phase change (BPSK, Fig. 2d). To the best of our knowledge, this is the first demonstration of BASK using graphene circuit, while previous works have only shown BPSK and BFSK[24,28,29]. By adding the BASK scheme, the three basic binary schemes have been completed using flexible graphene circuits. We note that the output voltage has a DC component for 0 and 1 because the transistor is operating at different bias points on the gate response curve. The DC component can be filtered out using a high pass filter and it has been removed for clarity in this paper.

Furthermore, by combining BASK and BPSK, a quaternary amplitude-shift keying (4-ASK) was demonstrated as shown in Fig. 2e. The inset of Fig. 2e illustrates the four bias points used in the 4-ASK scheme that correspond to 00, 01, 10, and 11. Both the phase and the amplitude information are used to distinguish the quaternary signal that is encoded in the carrier wave. Output of 00, 01, 10, and 11 are represented by "low amplitude, 270° phase", "high amplitude, 270° phase", "high amplitude, 90° phase", and "low amplitude, 90° phase" in the carrier wave, respectively. 4-ASK is a quaternary digital modulation scheme that uses four points in the constellation diagram (Fig. 1a) and doubles the data transfer rate compared to a binary scheme. Importantly, this is the first demonstration of quaternary modulation with just one transistor (excluding the transistor that is used as a resistor), which is not possible in conventional silicon based modulators.

**Quadrature phase-shift keying with two graphene transistors.**

A more fundamental quaternary modulation scheme is QPSK, which explores all four quadrants of the constellation. Figure 3a shows a typical QPSK transmitter structure used in modern digital communication. A binary data stream is demultiplexed into the in-phase component (I) and the quadrature-phase component (Q). I and Q components are encoded onto two orthogonal basis functions, such as a sine wave and a cosine wave, respectively, before they are summed to generate a QPSK modulated signal.

Here, we used just two transistors with similar gate response in the all-graphene circuit to demonstrate the QPSK modulation (Fig. 3b). Actual microscope images under a blue filter is overlaid on top of the circuit diagram. A sinusoidal wave from the function generator was connected to a simple off-chip resistance-capacitance – capacitance-resistance (RC-CR) phase shift network to generate two orthogonal wave functions with 90° phase difference. The sinusoidal input is shifted by +45° in the CR branch and by -45° in the RC branch [34]. Then each of these signals is summed internally by the function generators with two square waves ($I_{data}$ and $Q_{data}$) and fed to the gates of each transistor (A detailed measurement setup is shown in the Supplementary Fig. S2). The outputs ($V_I$ and $V_Q$) are then summed to generate the final QPSK modulated signals. These signal components are plotted in Fig. 3c. The $I_{carrier}$ and the $Q_{carrier}$ are the orthogonal carrier signals. The data bitstream with 00, 01, 10, 11 is represented by the in-phase component $I_{data}$ and the quadrature-phase component $Q_{data}$ as shown in the plot. Modulating $I_{carrier}$ with $I_{data}$ results in phase changes in $I_{channel}$ and the same applies to $Q_{carrier}$, $Q_{data}$, and $Q_{channel}$. Data bit 0 and 1 in $I_{data}$ corresponds to phase of 180° and 0° in $I_{channel}$. Similarly, Data bit 0 and 1 of $Q_{data}$ corresponds to phase of 90° and 270° in $Q_{channel}$. The sum of $I_{channel}$ and $Q_{channel}$ is the final output signal (I+Q) which has distinct phase shifts of 225°, 135°, 315°, and 45°, each corresponding to binary data of 00, 01, 10, and 11.

To validate the result, the instantaneous phase information was extracted from the final output signal (I+Q) and plotted as demodulated phase (Fig. 3c, bottom panel). This mathematical form of demodulation was achieved by extracting the phase information from the Hilbert transform of the output signal (I+Q). The plot of the demodulated phase indicates a clear distinction of phase shift between different QPSK signals. The carrier to noise ratio (C/N) which is the ratio of signal power to the white-noise power was found to be 21.1 dB from the frequency spectrum using a conventional signal analysis program (see Supplementary Methods). The corresponding bit error rate (BER) assuming additive white Gaussian noise (AWGN) channel is much lower than the performance threshold

BER of $10^{-4}$, above which the radio link is considered to be in outage [35]. This confirms the robustness and accuracy of the graphene based QPSK modulator. From the output signal ($(I+Q)$) and the input carrier signals, the gain of the QPSK modulator is calculated to be around 0.06, one of largest ever measured with graphene modulators. The gains of our all-graphene binary and quaternary modulators are comparable or larger than all the previous modulator works as shown in the Supplementary Table S1. Although these gain values are less than 1, in a modern transmitter structure, the amplification of the signal is primarily accomplished by an audio amplifier or a power amplifier, and a gain is not an essential component of the modulator.

**All-graphene modulator circuits under mechanical strain.**

Last, we examine the performance of all-graphene circuit under mechanical strain (Fig. 4). Frequency modulation (i.e., frequency doubling, Supplementary Fig. S3) was first evaluated at different bending radii. To quantify the comparison, fast Fourier transform (FFT) was applied to the doubled output voltage, yielding peaks at *ω* and *2ω* corresponding to the original frequency and the doubled frequency (Supplementary Fig. S4). The ratios of these two peaks, which indicate the spectral purity of the frequency doubling, are plotted as a function of the bending radii in Fig. 4a. Very little change is observed under different bending radii, indicating the robustness in circuit performance under mechanical strain. In addition, various binary and quaternary digital modulation schemes are also successfully demonstrated at maximum strain level of 2.7% (bending radius of 5.5mm) under the test set-up (Fig. 4b). The results further confirm the transparent all-graphene modulators are fully functional under highly strained conditions.

## Discussion

The operating principle and technique of flexible and transparent all-graphene modulators described here can be applied to widely used network technologies in today's multimedia and communication devices by either introducing pulse shaping with signal delays or coupling aforementioned modulation schemes [32]. Several recent works also demonstrated voltage gain in graphene transistors showing possibility of graphene based amplifiers and RF front-ends in radiofrequency communication system[36,37,38]. The combination of an efficient modulation method with a reliable RF front-end will be the key factor in determining the practicality of mobile and flexible apparatus. In conjunction with conventional thin film technology and high resolution lithography, all-graphene modulator circuit will play a pivotal role in realizing a high speed, mechanically compliant, and transparent electronic system in the near future.

## Methods

### Sample fabrication.

Graphene films used in this work are synthesized using chemcal vapor deposition (CVD) method on copper foil [39,40,41]. After the CVD synthesis, one side of the copper sample with graphene was coated with 950PMMA A2 (Microchem) resist and cured at 180°C for 1 minutes. The other side of the sample was exposed to $O_2$ plasma for 30 seconds to remove the graphene on that side. The sample was then left in Ammonium persulfate (Sigma Aldrich, 248614-500G) solution (0.025g/ml) for at least 6 hours to completely dissolve away the copper layer. Then the graphene was transferred to a 100um thick polyethylene naphthalate (PEN) substrate. The PMMA coating is removed with acetone and the substrate is rinsed with deionised water several times. The graphene layer was then patterned with a conventional stepper tool (GCA AS200 AutoStepper) using SPR220 3.0 (Microchem) resist. The process temperature was kept under the glass transition temperature of the plastic substrate (120°C) at all times. After graphene was patterned, 2nm of $Al_2O_3$ was deposited as a buffer layer using e-beam evaporation. Then 65nm of $Al_2O_3$ was deposited as the dielectric using atomic layer deposition at 80°C. Another graphene layer was transferred on top of the $Al_2O_3$ layer to form the channel layer and then it was patterned with lithography again. E-beam evaporation and atomic layer deposition of $Al_2O_3$ with the same thickness as the bottom dielectric was repeated on top of the channel layer. Final graphene layer was transferred again on top of the dielectric and patterned with lithography to be used as the top gate.

**Acknowledgements** Acknowledgement is made to the National Science Foundation Scalable Nanomanufacturing Program (DMR-1120187). Part of the work was conducted in the Lurie Nanofabrication Facility at University of Michigan, a member of the National Nanotechnology Infrastructure Network funded by the National Science Foundation.

**Author contributions** S.L. and Z.Z. conceived the experiments. S.L. fabricated the devices, developed the electrical measurement set-up and performed the measurements. K.L. provided support for fabrication, C.L. provided support for transmittance measurement, and G.S.K contributed to the electrical measurement set-up. S.L. wrote the paper and Z.Z. supervised the work. All authors discussed the results and commented on the manuscript.

**Additional information** The authors declare that they have no competing financial interest. Supplementary information accompanies this paper at www.nature.com/naturecommunications. Reprints and permission information is available online at http://npg.nature.com/reprintsandpermissions/. Correspondence and requests for materials should be addressed to Z.Z. (zzhong@umich.edu).

**Figure 1**

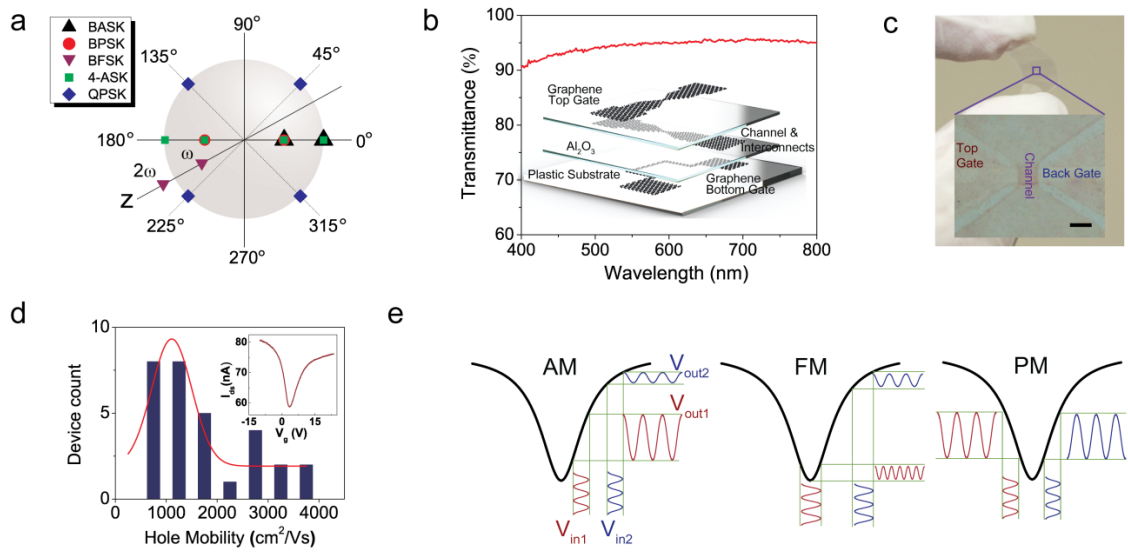

**Figure 2**

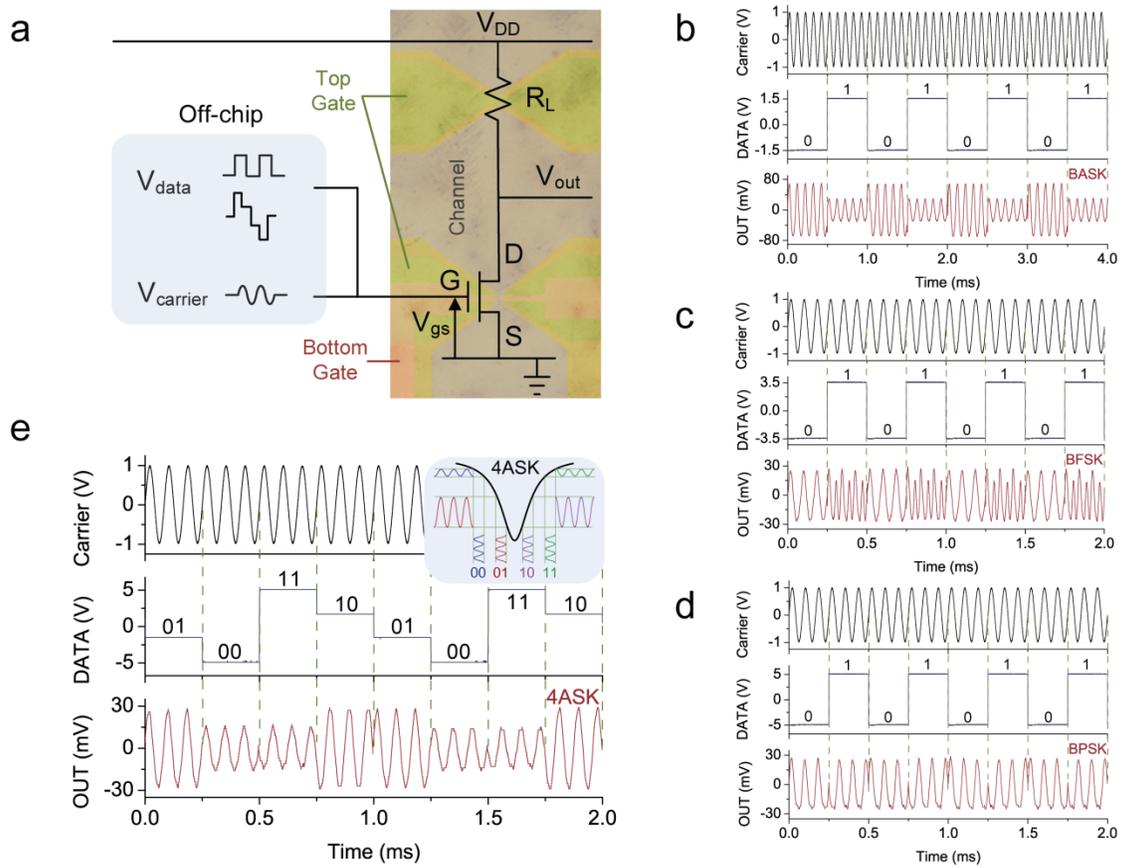

**Figure 3**

**Figure 4**

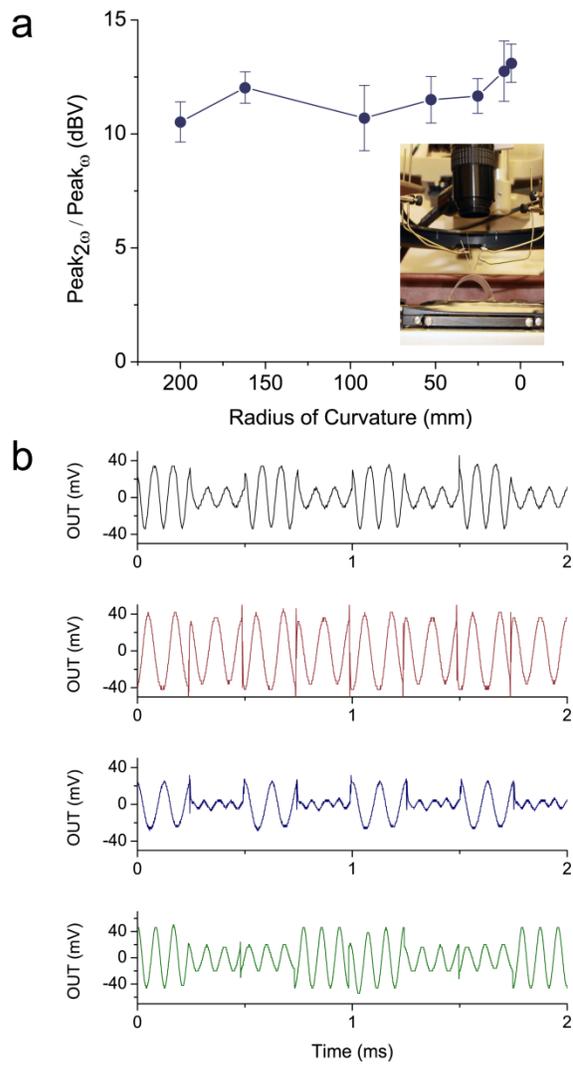

## Figure Captions

**Figure 1 Modulation mechanism and device characteristics of flexible and transparent all-graphene transistors. a,** A constellation diagram depicting five different digital modulation techniques demonstrated in this work. The z-axis, representing the frequency, is included to show the frequency modulated signals. **b,** A plot of the transmittance as a function of the wavelength and an illustration of the all-graphene transistor structure (inset). **c**, a photograph of graphene circuit on a transparent and bendable plastic substrate, and a microscopic image of an all-graphene transistor (inset). The scale bar is 10μm. **d,** A histogram of the hole mobility extracted from 30 transistors and its Gaussian fit (red line). The inset is a plot of the ambipolar current as a function of gate voltage for a typical all-graphene transistor. Voltage across the drain and the source is 10mV. **e,** Illustrations of amplitude, frequency, and phase modulation of a sinusoidal wave achieved by operating a single ambipolar graphene transistor at different gate biases.

**Figure 2 Binary and quaternary digital modulations using a single all-graphene transistor. a,** A circuit diagram with a false-color image of graphene transistors connected in a common-source configuration. The $V_{data}$ signal is the digital data that is encoded onto the carrier signal $V_{carrier}$. The $V_{data}$ signal is a square wave for all three binary digital modulation schemes and a four level step-like wave for the quaternary amplitude-shift keying modulation scheme. **b,** Time domain plot of the binary amplitude-shift keying. $V_{DD}$ of 1V was the power supply voltage. **c,** Time domain plot of the binary frequency-shift keying at $V_{DD}$ of 1V. **d,** Time domain plot of the binary phase-

shift keying at $V_{DD}$ of 1V. **e**, A time domain plot of the quaternary amplitude-shift keying modulation of carrier signal. The inset is an illustration describing the four operating gate bias points used in 4-ASK. $V_{DD}$ is 1V.

**Figure 3 Quadrature phase-shift keying demonstrated with two all-graphene transistors. a,** A conceptual diagram of a conventional quadrature phase-shift keying transmitter structure. NRZ encoder is a non-return-to-zero encoder where 1 is represented by a positive voltage state and 0 is represented by a negative voltage state. RC-CR network is the resistance-capacitance–capacitance-resistance phase shift network which generates two orthogonal wave functions with 90° phase difference. **b,** An all-graphene circuit diagram of the quadrature phase-shift keying system using two transistors. The actual microscopic image of the all-graphene circuit under a blue filter is shown. The transistor dimension is 10μm × 10μm. **c,** Time domain plots of the input and output signals demonstrating quadrature phase-shift keying modulation scheme. $V_{DD}$ of 7V was the power supply voltage.

**Figure 4 Flexible and transparent all-graphene digital modulator circuits under mechanical strain. a,** The plot of the signal amplitude ratio of the original and the doubled frequency as a function of the curvature radius for a graphene frequency doubler. The inset is a photograph of the measurement setup. **b,** Time domain plots of binary amplitude-shift keying (black), binary phase-shift keying (red), binary frequency-shift keying (blue), and quaternary amplitude-shift keying (green) schemes achieved with mechanically strained all-graphene circuits at 5.5mm radius of curvature (2.7% strain).

Supplementary Information for

# Flexible and Transparent All-Graphene Circuits for Quaternary Digital Modulations

Seunghyun Lee[1], Kyunghoon Lee[1], Chang-Hua Liu[1], Girish S. Kulkarni[1] and Zhaohui Zhong[1]*

[1] Department of Electrical Engineering and Computer Science, University of Michigan, Ann Arbor, MI 48109, USA.

* e-mail: zzhong@umich.edu

Supplementary Figures

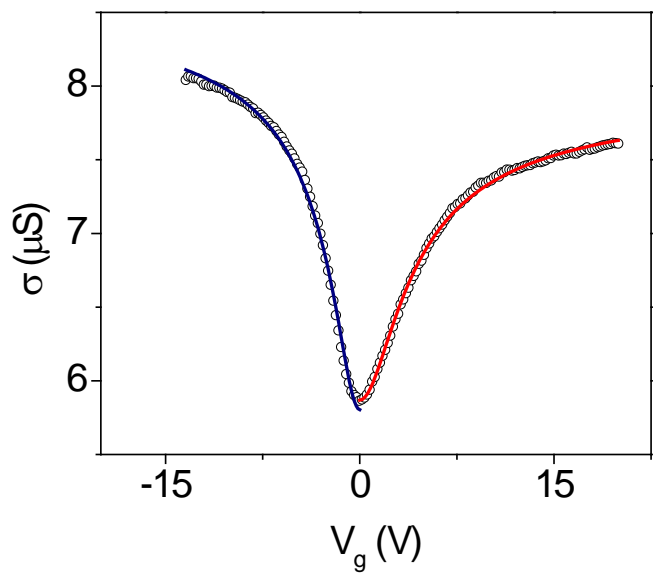

**Supplementary Figure S1, Conductance as a function of gate voltage (round symbols) and its fittings (solid line) for a typical device.** The blue and the red solid lines correspond to the fittings for hole and electron mobility respectively. See Supplementary Methods for details on the fitting method.

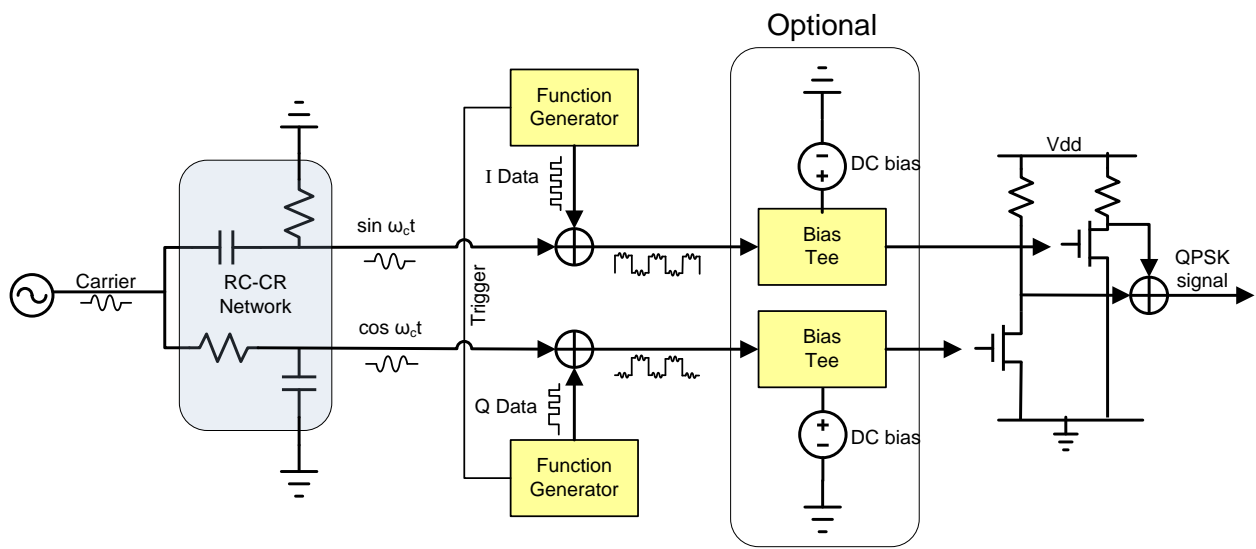

**Supplementary Figure S2, Detailed measurement setup for quadrature phase shift-keying signal generation.** A sinusoidal carrier signal is generated from a signal generator and fed to an RC-CR phase shift network. The phase of the sinusoidal signal is shifted by +45 and -45 when it passes through RC and CR structure respectively. The resulting two orthogonal functions ($\sin \omega_c t$ and $\cos \omega_c t$) with a phase difference of 90° are summed internally in two different function generators with its respective digital data signal shown as the square wave. The two function generators are phase matched using the trigger function. If the charge neutrality point ($V_{Dirac}$) is not centered at zero voltage due to environmental doping, the signal can be connected with a bias tee with a DC bias and then fed to the gate of each transistor. When the DC bias is approximately equal to $V_{Dirac}$, the phase modulation of each transistors will be symmetric. If the Dirac point at 0 voltage, the signals can be directly inserted to the gates of each transistors without a bias tee The two generated signals which are the final quadrature phase-shift keying signals were added internally and measured with an oscilloscope.

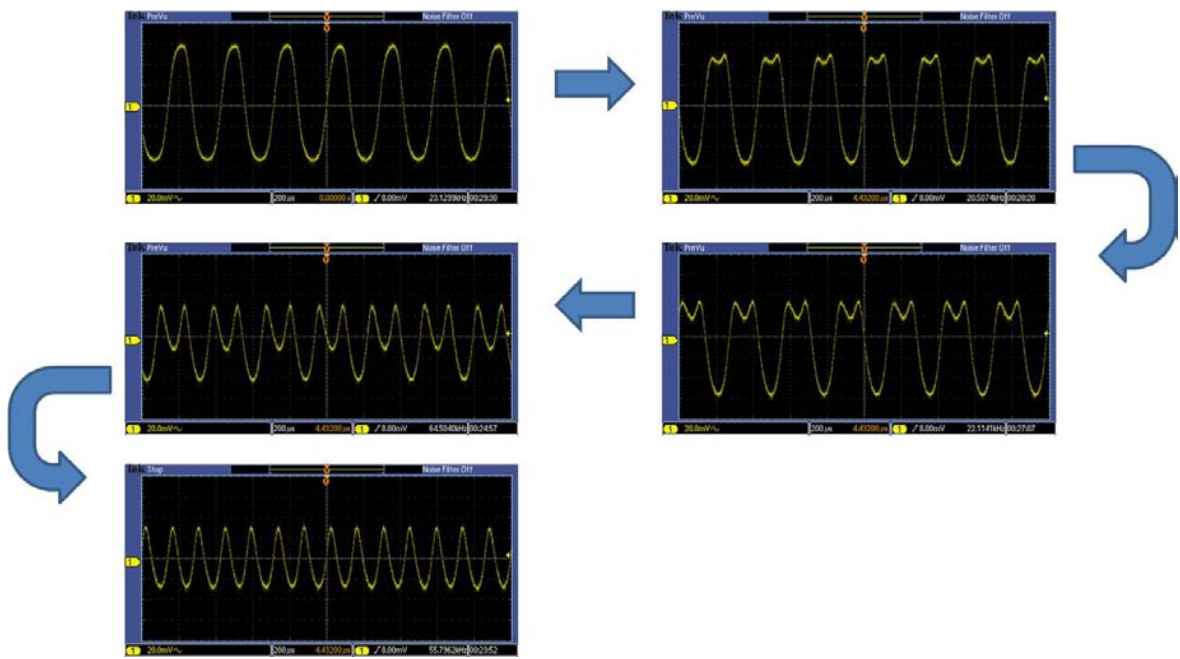

**Supplementary Figure S3, Oscillscope images of the frequency doubling as a result of gradual gate bias shift.** As the gate DC bias point shifts from the negative side (hole carrier dominated) to the Dirac point, frequency doubling due to ambipolar characteristics of graphene transistor can be observed. If the DC bias point is not exactly at the Dirac point, the output signal will show asymmetry.

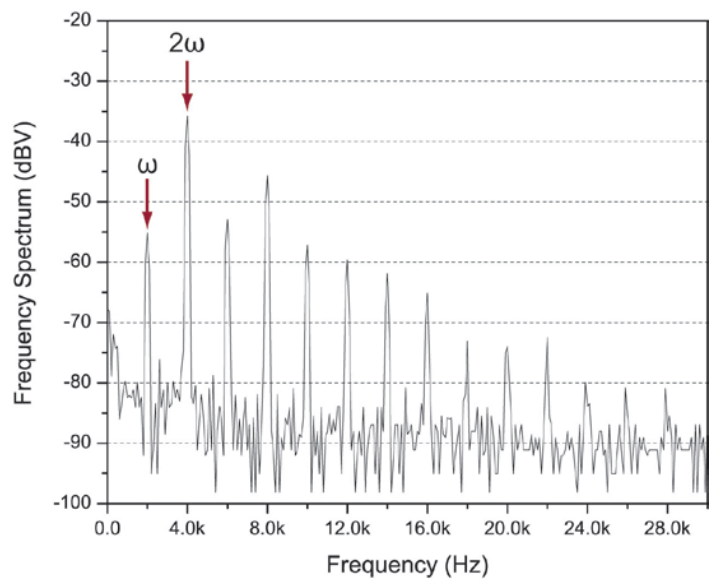

**Supplementary Figure S4. Fast Fourier Transform (FFT) of a typical frequency doubled signal.** A typical Fast Fourier Transform (FFT) of the doubled signal from a mechanically strained graphene transistor is shown. The radius of curvature was 5.5mm for this plot. The doubled frequency ($2\omega$) term and the original frequency ($\omega$) term has a signal amplitude difference of 20 dBV. The higher order terms also shows significantly weaker signal strength compared to the doubled frequency. The higher order terms can be filtered out if necessary.

## Supplementary Table

| Publication | Gain | Modulation frequency | Comments |
|---|---|---|---|
| Wang et.al.[25] | 0.005 | 10 kHz | Frequency doubling |
| Yang et. al.[24] | ~0.01 | 4-10 kHz | BPSK, BFSK |
| Hsu et. al.[29] | 0.005 | 500 Hz | BPSK |
| Harada et.al.[28] | ~0.05 | 30 kHz | BPSK |
| Sordan et.al.[42] | <0.025 | 100 Hz | Boolean logic |
| **This work : binary modulation** | **0.03~0.07.** | **10 kHz** | **BPSK, BFSK, BASK. The first demonstration of BASK.** |
| **This work : quaternary modulation** | **4ASK : 0.03  QPSK : 0.06** | **10 kHz** | **4-ASK, QPSK. The first demonstration of quaternary modulations.** |

**Supplementary Table S1, Signal gain comparison of past works and this work.**

## Supplementary Methods

**Transmittance measurement**

The transmittance measurement setup consists of a monochromator (Acton SP2300 triple grating monochromator/spectrograph, Princeton Instruments) coupled with a 250W tungsten halogen lamp (Hamatsu), a collimator, and a photodetector. An iris was used to prevent the photodetector from absorbing the scattered light from the substrate. Optical power measurements were carried out using a 1928-C power meter (Newport) coupled to a UV enhanced 918UV Si photodetector (Newport). A blank PEN substrate was used as a reference for subtraction.

**DC characterization**

The contact resistance and the mobility can be extracted by fitting the experimental value of resistance across the source and drain of the graphene transistors with the following equation[23],

$$R_{total} = \frac{V_{ds}}{I_{ds}} = R_{contact} + \frac{L}{q\mu W \sqrt{n_o^2 + (C_{ox}\frac{(V_g - V_{Dirac})}{q})^2}}$$

(Supplementary Equations S1)

where the variables are defined as drain/sourse voltage $V_{ds}$, drain/sourse current $I_{ds}$, contact resistance $R_{contact}$, gate capacitance $C_{ox}$, residual carrier concentration $n_o$, the gate voltage $V_g$, the charge neutrality point $V_{Dirac}$, drain/source width $W$ and length $L$. Device in Supplementary Fig. S1 indicated a hole mobility of 3342±26 cm$^2$/Vs and electron mobility of 2813±11 cm$^2$/Vs with residual concentration of $n_o$= (2.47±0.01)×10$^{11}$ cm$^{-2}$, and $R_{contact}$ = 116.4±0.1 kΩ. For all the cases, the residual concentration matched well with the reported values 2×10$^{11}$ cm$^{-2}$ [43]. Notably, the high contact resistance is resulted from series resistance of long graphene strips which have been used as the interconnects between the drain/source electrodes and the contacts. Although the large series resistance currently limits the frequency performance of the devices, this problem can be resolved by partial doping of graphene interconnects in the future. Several works have shown it is possible to lower the graphene sheet resistance significantly by room temperature doping[41,44].

**Extraction of carrier to noise ratio (C/N)**

Signal to noise ratio is defined as,

$$10\log_{10} \frac{V_{Signal}^2}{V_{Noise}^2} \text{ [dB]}$$

(Supplementary Equations S2)

This figure characterizes the ratio of the fundamental signal to the noise spectrum. The noise spectrum includes all non-fundamental spectral components such as spurs and the noise floor in the Nyquist frequency range (sampling frequency / 2) without the DC component, the fundamental itself and the harmonics. Six harmonics were considered in our calculation. Carrier to noise ratio (C/N) [i.e. signal to noise ratio of a modulated signal] of 21.1 dB was extracted from the fast Fourier transform plot using a conventional program, SBench 6.1 (Spectrum GmbH). The bit error rate (BER) from this C/N value is significantly better than the performance threshold of a QPSK system as indicated in the reference[45].

# Supplementary References